\documentstyle[aps,amssymb,psfig,graphics]{revtex}

\begin{document}
\title{Green fluorescent proteins as optically-controllable elements in 
bioelectronics}
\author{Riccardo A. G. Cinelli$^{1}$, Vittorio Pellegrini$^{1,}$$^{a)}$, 
Aldo Ferrari$^{1}$, Paolo Faraci$^{1}$, Riccardo Nifos\`\i$^{1}$, Mudit 
Tyagi$^{1,2}$, Mauro Giacca$^{1,2}$, and Fabio Beltram$^{1}$}
\address{1. NEST-INFM and Scuola Normale Superiore, Piazza dei Cavalieri 7, 
I-56126 
Pisa, Italy}
\address{2. Molecular Medicine Laboratory, International Centre for Genetic 
Engineering
and Biotechnology, Padriciano 99, I-34012 Trieste, Italy}
\footnotetext[1]{E-mail: vp@sns.it}
\maketitle
\begin{abstract}
A single-biomolecule optical toggle-switch is demonstrated based on a mutated 
green fluorescent protein (GFP). We have exploited molecular biology 
techniques to tailor GFP molecular 
structure and photophysical properties and give it optically-controlled 
bistability between 
two distinct states. 
We present optical control of the fluorescence dynamics 
with two laser beams at 476 and 350 nm down to the ultimate limit of single 
molecules. 
These results indicate that GFP-class fluorophores are promising candidates 
for the realization of 
biomolecular devices such as volumetric optical memories and optical switches.

\end{abstract}
\newpage

The optical control of the molecular structure of materials is an attractive 
way to store and 
manipulate data at high spatial 
resolution\,\cite{libro_memory}. Promising candidates for this application are 
photochromic molecules that exhibit photoinducible isomerization between forms 
with different 
optical properties. 
The potential of photochromism has generated a large 
interest in the synthesis of new photosensitive materials and in the 
development 
of new techniques for their use\,\cite{photochromism}. In this framework, some 
biological systems, like bacteriorhodopsin and the green fluorescent 
protein (GFP) of the {\it Aequorea victoria} jellyfish, are offering 
naturally-evolved optimized structures with unique properties that can 
be further tailored for specific applications by genetic 
engineering\,\cite{rhodopsin,gfpblinking}. 
While bacteriorhodopsin has already been the subject of an intense 
research effort and was proposed for a broad range of bioelectronic 
applications\,\cite{birge}, GFP potential for bioelectronics is still 
largely unexplored in spite of the advantage of a very efficient 
fluorescence emission that allows detection down to the single-molecule 
level\,\cite{gfpblinking,cinelli,pierce}.

GFP has emerged in recent years as a unique fluorescent marker in molecular 
and cell biology\,\cite{gfplibro}. Molecular engineering is playing a 
fundamental role in designing GFP mutants with modified spectral 
characteristics, enhanced brightness, photostability, quantum yield, 
and other properties tailored for different 
applications. GFP fluorescence dynamics is characterized by transitions 
between bright and dark states which, at the single-molecule level, leads to
reversible turning on and off (blinking) and ultimate switching off 
(photobleaching)
of the emission\,\cite{gfpblinking,cinelli,pierce}. 

The availability of distinct ground states is a requirement for 
the development of high-density volumetric optical memories and optical 
switches. 
Other important issues, however, must be addressed such as efficient control 
of the photoconversion between these different states and achievement of high 
spatial resolution.
A first important step in this direction was the observation of 
photobleaching reversal in two yellow-shifted GFP mutants by 
means of prolonged (5 min) irradiation at 405 nm by a Hg arc 
lamp\,\cite{gfpblinking}. Control of photoconversion between dark and bright 
states is not achievable with other GFP mutants like the F64L/S65T GFP 
(enhanced GFP, EGFP), one of the most common optical markers in 
biology\,\cite{cinelli}. 

In this letter we demonstrate that one specific point mutation (T203Y, 
threonine 
into tyrosine at position 203 in the amino-acid sequence) enables
the optical control of transitions between 
two distinct states down to the ultimate limit of single molecules. 
We show that it is possible to induce unlimited optically-controllable cycles 
between fluorescent and dark configurations by means of two laser beams 
at the wavelengths of 476 and 350 nm. 
Additionally a model for the observed photophysics is suggested based on the 
selectivity of photoconversion on the excitation wavelength.

EGFP and T203Y EGFP (E$^2$GFP in the following) molecules \cite{preparazione} 
were obtained as recombinant proteins and studied both in solution and trapped 
in polyacrylamide (PAA) gel films ($\approx$10 $\mu$m thick)\,\cite{gel}. 
Gel preparation provided pore sizes small enough for immobilization of 
proteins while maintaining their native conformation\,\cite{cinelli}. 
At sufficiently low fluorophore concentration (around 5 ng/cm$^2$), optical 
emission from isolated single molecules and cluster could be observed. 
Clusters originate from unavoidable inhomogeneities  of the PAA gel that  
act as localization centers for the fluorophores\,\cite{cinelli}.
By comparing fluorescence intensities of single molecules (see below) and 
clusters, we could establish that clusters are typically composed of 
approximately one hundred molecules. For both molecules, room-temperature 
fluorescence was excited by an Argon-ion laser at 476 nm. Radiation at 406 nm 
and 350 nm was provided by an Argon/Krypton-ion laser. The two lasers were 
focused onto a circular field of view of $\sim$10 $\mu$m$^2$ in PAA gels by a 
100$\times$ 1.3 N.A. oil-immersion objective and GFP molecules were imaged by 
an 
intensified CCD camera (0.1 s integration time).

E$^2$GFP absorption measurements were performed on 300 $\mu $l solution (70 
$\mu$M) 
prepared from PBS (phospate buffered saline) at pH=7.0
and display an equilibrium between two states (see Fig. 1 upper panel, dotted 
line)
analogous to that of wild-type GFP\,\cite{kummer_yellow}. 
The absorption peaks at 400 and 515 nm suggest that these states (A and B) 
are associated to the protonated neutral and deprotonated 
anionic form of the chromophore, respectively\,\cite{brejc}. 
On the contrary, EGFP shows a very different 
equilibrium strongly favoring state B (see Fig.1 lower panel, dotted line). 
Fluorescence after excitation of state B 
is shown in Fig.1 for both mutants (solid lines). In the case of E$^2$GFP, 
fluorescence peaks at 
523 nm and is red-shifted with respect to EGFP (512 nm).
Excitation of state A in both molecules produces a similar but weaker 
fluorescence 
signal which was interpreted in wild-type GFP in terms of excited-state 
photoconversion from A to B through a proton-transfer 
process\,\cite{chattorai}.

Single-molecule studies were performed in gel-trapped samples. 
We observed blinking and photobleaching into a long-lasting dark state 
(state C in the following) within few seconds after excitation at 476 nm. 
The top histogram of Fig. 2, however, shows that photoconversion from state 
C back into state B of E$^2$GFP is possible and
can be driven by irradiation at 350 nm. The histogram reports 
the fluorescence intensity of an E$^2$GFP cluster in PAA gel composed of 
approximately 50 molecules after alternate 10 s-long excitation at 
476 nm and 2 s at 350 nm (0.2 kW/cm$^2$ at both wavelengths). 
The measurements are an extract of an arbitrarily-extendable experiment and 
show 
photobleaching by excitation at 476 nm and efficient photoconversion from C to 
B 
by irradiation at 350 nm. 
In order to rule out possible changes in the ground-state configuration after 
excitation 
in the UV, we also analyzed the E$^2$GFP emission lineshapes after 
photoconversion. 
The lineshapes (data not shown) are identical to that reported in Fig.1 
confirming 
that the 350 nm irradiation indeed converts the protein back to the original 
state B. 
In addition, we noted selectivity of photoconversion on wavelength: radiation 
at 406 nm (from 0.1 to 10 kW/cm$^2$) instead of 350 nm favored E$^2$GFP 
photobleaching indicating that state C is distinct from state A. 
The bottom histogram shown in Fig. 2 presents the analogous experiment for an 
EGFP cluster. Following the initial photobleaching, no significant recovery is 
observed. This shows unambiguously that photoconversion originates from the 
single point-mutation T203Y. 

States B and C thus encode a (0, 1) bit that can be stored and manipulated in 
the protein. 
However, in a memory device we must be able to write, read and erase data. 
These operations can be accomplished with E$^2$GFP. One possible 
implementation is 
to exploit photoconversion from C to B with irradiation at 350 nm ({\tiny 
{WRITE}}), 
fluorescence emission following weak excitation at 476 nm ({\tiny {READ}}) 
and photobleaching (B to C) ({\tiny {ERASE}}). 
The last process can be induced  by intense or prolonged excitation at 476 nm.

One important issue for advanced bioelectronic applications
concerns the possibility to address the protein state at the single-molecule 
level. 
To this end we investigated the limit for the observation of E$^2$GFP 
optical switching presented above. Individual E$^2$GFP molecules 
in state B were imaged as bright spots at the resolution limit of the 
optical setup ($\sim$0.3 $\mu$m). The fluorescence intensity of single spots 
($\sim$300 CCD counts) 
was similar to EGFP and displayed the characteristic signatures of 
single-molecule emission. 
Most of the molecules, in fact, blinked and then photobleached within 10 s 
under excitation at 476 nm. 
However, compared to EGFP, we typically observed longer duration of E$^2$GFP 
emission. 
Blinking events were detected at longer times (up to $\sim$1 min). 
Figure 3 shows that controlled photoconversion is achieved both on E$^2$GFP 
clusters and E$^2$GFP single molecules. In this latter case, the same molecule 
(three lower images) 
was repeatedly photobleached by excitation at 476 nm ({\tiny{ERASE}}), 
photoconverted with 350 nm laser irradiation 
({\tiny{WRITE}}) and its fluorescent emission detected with 100 ms integration 
time ({\small{READ}}). 
The three upper images of the figure show efficient photoconversion 
for a cluster of around one hundred molecules and confirm the behavior of the 
histograms shown in the top part of Fig. 2.

In order to rationalize the observed differences between E$^2$GFP and EGFP, we 
performed molecular dynamics simulations of the whole proteins using the AMBER 
suite of programs\,\cite{Amber}. Figure 1 shows our results
for anionic states B in the two mutants (chromophores and their immediate 
environment). 
Similarly to other T203Y mutants, tyrosine at site 203 interacts with the 
chromophore 
through $\pi$-stacking\,\cite{redshift}. In E$^2$GFP, this $\pi$-stacking and 
the peculiar 
hydrogen bonding 
around the chromophore phenolic ring  cause the red-shift and destabilize 
state B in favor of state A as observed in the absorption and emission spectra 
reported in 
Fig. 1\,\cite{redshift}. 
These simulations suggest that the peculiar hydrogen-bond network resulting 
from 
the point mutation at position 203 is responsible for the photoconvertibility 
between bright (B) and dark (C) states. Photobleaching and photoconversion 
most 
probably involve a change in the protonation state of the chromophore, i.e. a 
proton transfer. 
The inset in the top panel of Fig.2 
describes a working model of E$^2$GFP photophysics based on these results. 
The equilibrium between states A and B as schematically indicated by the 
arrows 
follows the results of Ref. \onlinecite{kummer_yellow}. Our data indicate that
we must include two additional distinct states responsible for blinking and 
photobleaching. 
The zwitterion was indicated as the chromophore form associated to the 
blinking 
dark state Z\,\cite{weber}. The latter reference also discusses the 
equilibrium between Z and B states. 
Additionally our experiments show photoconvertibility of E$^2$GFP from C to B 
following absorption 
at 350 nm (i.e. from the C* excited state to B) and from B to C following 
absorption at 
476 nm (i.e. from the B* excited state to C). We should like to underline that 
this situation 
differs from what observed in yellow-shifted GFP 
mutants\,\cite{gfpblinking,weber}. 
In the latter case photoconversion was reported between states A and B without 
invoking an 
additional chromophore state. 
Further analysis based on ab-initio calculations is required to 
identify the nature of C and the precise mechanism of conversion among the 
different configurations. We do expect that this analysis will make it 
possible 
to suggest new mutations able to further tune the potential barriers 
separating different states.

In conclusion, we demonstrated that a single point mutation T203Y of the 
enhanced green fluorescent protein yields a photochromic behavior and an 
optically controlled molecular toggle-switch by means of two focused laser 
beams at different wavelengths. 
We showed that controlled and efficient photoconversion between two distinct 
ground states is possible down to the ultimate limit of the single molecule. 
We argue that this process can represent the basis for the implementation 
of dense volumetric GFP-based optical memories that exploit the fluorescent 
properties of single proteins. Finally we note that the photoconversion of 
E$^2$GFP opens the way to advanced biological applications where prolonged 
monitoring of GFP-tagged molecules in live cells is required. This may lead to 
a new level of resolution in the study of biomolecular processes.

{\bf Acknowledgements}. Work at Scuola Normale Superiore was funded in part by 
MURST. We thank Valentina Tozzini for useful discussions.

\newpage
\begin{figure}
\caption{Normalized absorption (dotted lines) and emission (following 
excitation at 476 nm, 
solid lines) spectra of E$^2$GFP (upper panel) and EGFP (lower panel).
Structures of the chromophore environment of E$^2$GFP (top) and EGFP (bottom) 
in the anionic B state 
following $\sim$1 ns molecular dynamics simulations are 
also reported. Dashed lines indicate hydrogen bonds. The hydrogen atoms are 
hidden and 
the water molecules are shown as balls. The picture was produced by 
Insight2000 
(Molecular Simulations Inc.).}
\end{figure}

\begin{figure}
\caption{Fluorescence intensities of E$^2$GFP (first 
histogram) and EGFP (second histogram) clusters in PAA gels after alternate 10 
s-long 
excitation at 476 nm and 2 s at 350 nm (0.2 kW/cm$^2$ at both wavelengths). 
Horizontal dots are guides to the eye at the same background level. 
Model for E$^2$GFP photophysics is described in the top panel.
On the left, the E$^2$GFP bright states A 
and B (associated to the neutral and the anionic chromophore forms, 
respectively) 
and their excited states A* and B* are indicated together with their 
excitation and 
emission wavelengths in nm. On the right, E$^2$GFP dark states Z and 
C are reported. Transitions between bright and dark states in blinking, 
photobleaching, 
and photoconversion are shown. Excitation from C to C* at 350 nm is deduced 
from 
the selectivity of photoconversion on the wavelength.}
\end{figure}

\begin{figure}
\caption{Typical images (from left to right) of an E$^2$GFP cluster (upper 
row) and a single 
E$^2$GFP molecule (lower row) in PAA gels after alternate photobleaching by 10 
s 
excitation at 476 nm and photoconversion by 2 s at 350 nm (0.2 kW/cm$^2$ at 
both wavelengths). 
Each frame size is 2 $\mu$m $\times$ 2 $\mu$m and integration time 100 ms}
\end{figure}

{\centerline{\psfig{figure=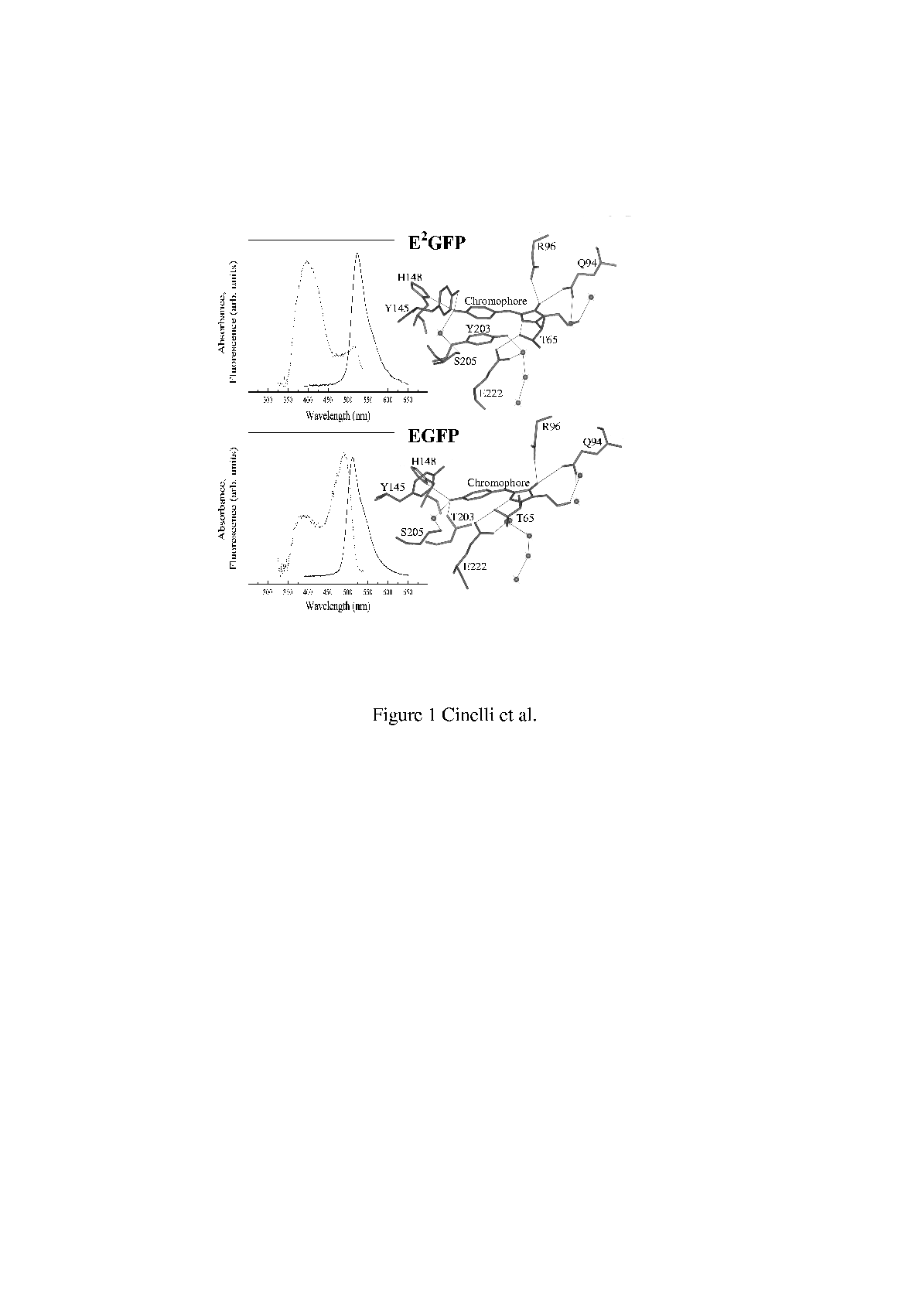,width=\linewidth}}}
{\centerline{\psfig{figure=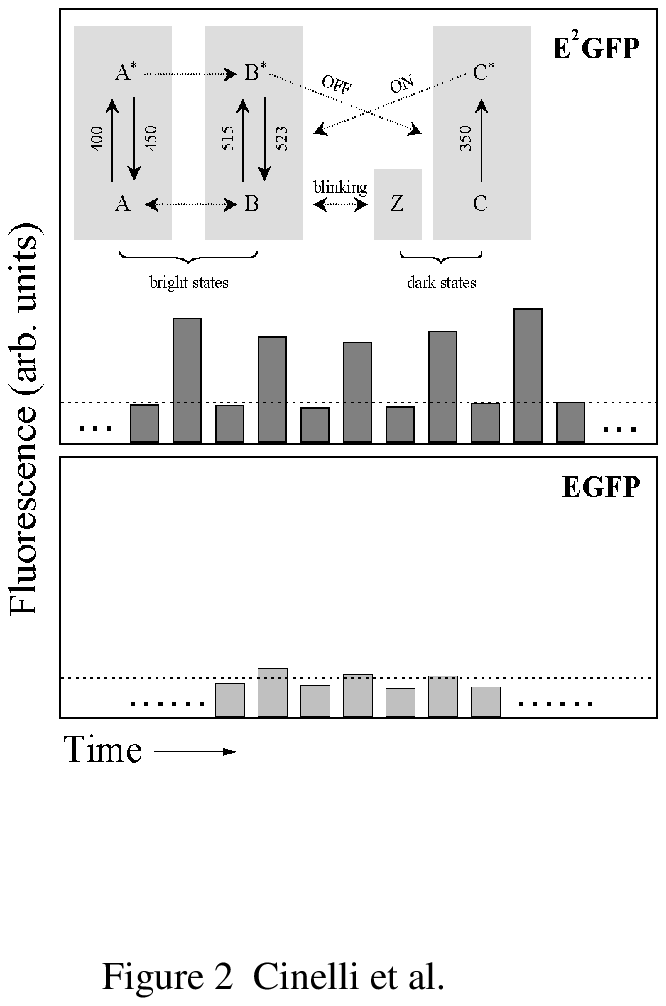,width=0.5\linewidth}}}
\centerline{\psfig{figure=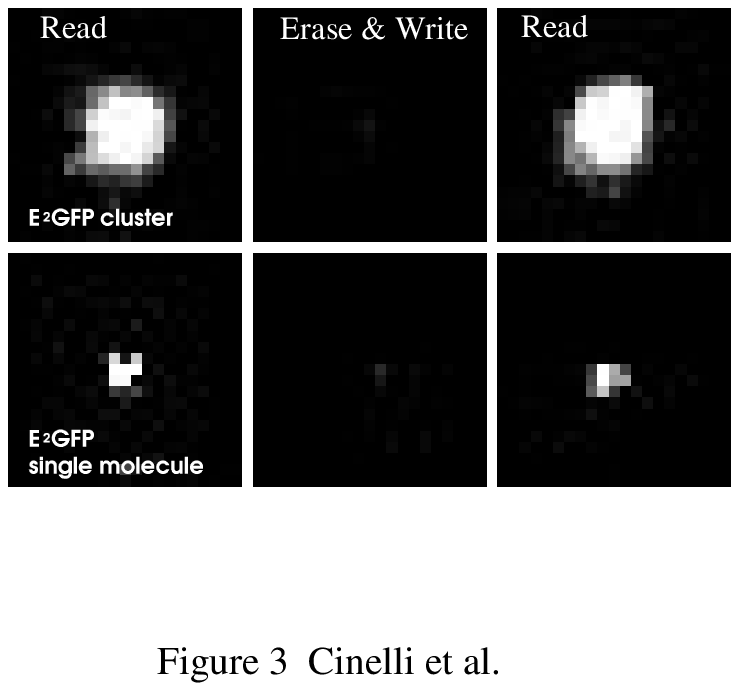,width=\linewidth}}

\newpage 
\noindent 

\begin{references}

\bibitem{libro_memory}
{\em Photo-Reactive Materials for Ultrahigh Density Optical Memory},
edited by M. Irie
(Elsevier, Amsterdam, 1994).

\bibitem{photochromism}
A. Toriumi, J. M. Herrmann, and S. Kawata,
{\em Opt. Lett.} {\bf 22}, 555 (1997);
K. Sasaki and T. Nagamura,
{\em Appl. Phys. Lett.} {\bf 71}, 434 (1997);
T. Tsujioka, Y. Hamada, K. Shibata, A. Taniguchi, and T. Fuyuki,
{\em Appl. Phys. Lett.} {\bf 78}, 2282 (2001).

\bibitem{rhodopsin}
A. Lewis, Y. Albeck, Z. Lange, J. Benchowski, and G. Weizman,
{\em Science} {\bf 275}, 1462 (1997);
T. Zhang, C. Zhang, G. Fu, Y. Li, L. Gu, G. Zhang, Q. W. Song, B. Parsons, 
and R. R. Birge,
{\em Opt. Eng.} {\bf 39}, 527 (2000);
T. M. H. Creemers, A. J. Lock, V. Subramaniam, T. M. Jovin, and S. V\"{o}lker,
{\em Proc. Natl. Acad. Sci. USA} {\bf 97}, 2974 (2000).

\bibitem{gfpblinking}
R. M. Dickson, A. B. Cubitt, R. Y. Tsien, and W. E. Moerner,
{\em Nature} {\bf 388}, 355 (1997).

\bibitem{birge}
R. R. Birge, N. B. Gillespie, E. W. Izaguirre, A. Kusnetzow, A. F. Lawrence, 
D. 
Singh, Q. W. Song, E. Schmidt, J. A. Stuart, S. Seetharaman, and K. J. Wise,
{\em J. Phys. Chem. B} {\bf 103}, 10746 (1999).

\bibitem{cinelli}
R. A. G. Cinelli, A. Ferrari, V. Pellegrini, M. Tyagi, M. Giacca, and 
F. Beltram,
{\em Photochem. Photobiol.} {\bf 71}, 771 (2000).

\bibitem{pierce}
D. W. Pierce, N. Hom-Booher, and R. D. Vale,
{\em Nature} {\bf 388}, 338 (1997);
E. J. G. Peterman, S. Brasselet, and W. E. Moerner,
{\em J. Phys. Chem. A} {\bf 103}, 10553 (1999);
M. F. Garcia-Parajo, G. M. J. Segers-Nolten, J.-A. Veerman, J. Greve, and 
N. F. van Hulst,
{\em Proc. Natl. Acad. Sci. USA} {\bf 97}, 7237 (2000).

\bibitem{gfplibro}
{\em Green Fluorescent Proteins},
edited by K. F. Sullivan and S. A. Kay
(Academic Press, San Diego, 1999).




\bibitem{preparazione}
Green fluorescent proteins were obtained following the procedures described in 
Ref.6.
The presence of mutation at appropriate site (T203Y) was confirmed by 
nucleotide sequencing.

\bibitem{gel}
PAA gels (T=15\%, C=3\% without sodium 
dodecyl sulfate; catalysis: tetramethylendiamine and ammonium persulfate) were 
prepared in pH 7 phosphate buffered saline doped with the protein. T is the 
total concentration of monomer in g per 100 ml, C is the wt\% of total monomer 
which is N,N'-methylenebisacrylamide.

\bibitem{kummer_yellow}
A. D. Kummer, C. Kompa, H. Lossau, F. P\"{o}llinger-Dammer, M. E. 
Michel-Beyerle, 
C. M. Silva, E. J. Bylina, W. J. Coleman, M. M. Yang, and D. C. Youvan,
{\em Chem. Phys.} {\bf 237}, 183 (1998).

\bibitem{brejc}
K. Brejc, T. K. Sixma, P. A. Kitts, S. R. Kain, R. Y. Tsien, M. Orm\"{o},
and S. J. Remington,
{\em Proc. Natl. Acad. Sci. USA} {\bf 94}, 2306 (1997).

\bibitem{redshift}
M. Orm\"{o}, A. B. Cubitt, K. Kallio, L. A. Gross, R. Y. Tsien, and S. J. 
Remington,
{\em Science} {\bf 273}, 1392 (1996);
R. M. Wachter, M.-A. Elsliger, K. Kallio, G. T. Hanson, and S. J. Remington,
{\em Structure} {\bf 6}, 1267 (1998);
A. D. Kummer, J. Wiehler, H. Rehaber, C. Kompa, B. Steipe, and M. E. 
Michel-Beyerle,
{\em J. Phys. Chem. B} {\bf 104}, 4791 (2000).

\bibitem{chattorai}
M. Chattoraj, B.A. King, G.U. Bublitz, and S.G. Boxer, {\em Proc. Natl. Acad. 
Sci. USA} 
{\bf 93}, 8362 (1996). 

\bibitem{Amber}
W. D. Cornell, O. Cieplak, C. I. Bayly, I. R. Gould, K. M. Merz Jr., D. M. 
Ferguson, 
D. C. Spellmeyer, T. Fox, J. W. Caldwell, and P. A. Kollman,
{\em J. Am. Chem. Soc.} {\bf 117}, 5179 (1995).

\bibitem{weber}
W. Weber, V. Helms, J. A. McCammon, and P. W. Langhoff,
{\em Proc. Natl. Acad. Sci. USA} {\bf 96}, 6177 (1999).


\end{references}
\end{document}